\documentclass[11pt,draftclsnofoot,onecolumn]{IEEEtran}

\usepackage[dvips]{graphicx}
\usepackage{amssymb}
\usepackage[cmex10]{amsmath}
\usepackage[caption=false]{subfig}
\usepackage{multirow}
\usepackage{array}

\newtheorem{definition}{Definition}
\newtheorem{proposition}{Proposition}
\newtheorem{assumption}{Assumption}

\newtheorem{corollary}{Corollary}

\begin{document}

\title{Intervention Mechanism Design\\ for Networks With Selfish Users}

\author{Jaeok Park and Mihaela van der Schaar\thanks{The authors are with Electrical Engineering Department, University of
California, Los Angeles (UCLA), 420 Westwood Plaza,
Los Angeles, CA 90095-1594, USA. e-mail:
\{jaeok, mihaela\}@ee.ucla.edu.}}

\maketitle

\begin{abstract}
We consider a multi-user network where a network manager
and selfish users interact.
The network manager monitors the behavior of users and intervenes in the
interaction among users if necessary, while users make decisions independently
to optimize their individual objectives.
In this paper, we develop a framework of intervention mechanism
design, which is aimed to optimize the objective of the manager, or the network
performance, taking the incentives of selfish users into account.
Our framework is general enough to cover a wide range of application scenarios,
and it has advantages over existing approaches such as Stackelberg strategies and pricing.
To design an intervention mechanism and to predict the resulting operating point,
we formulate a new class of games called intervention games
and a new solution concept called intervention equilibrium.
We provide analytic results about intervention equilibrium and
optimal intervention mechanisms in the case of a benevolent
manager with perfect monitoring. We illustrate these results
with a random access model. Our illustrative example suggests
that intervention requires less knowledge about users than pricing.
\end{abstract}


\section{Introduction}

\subsection{Motivation and Our Approach}

In noncooperative multi-user networks, multiple users interact with each other
making decisions independently to optimize their individual objectives.
Operation by selfish users often results in a suboptimal network performance,
and this calls for an incentive mechanism to guide selfish users
to behave according to the system objective.
In this paper, we propose a class of incentive mechanisms, called \emph{intervention
mechanisms}, that can be used when there is a network manager that can monitor
the behavior of users and intervene in the interaction among users if necessary.
Since the manager can react to the behavior of users through an intervention
mechanism, an intervention mechanism has the potential to shape the incentives
of users.

To design an intervention mechanism and to predict the resulting operating point,
we formulate a new class of games called \emph{intervention games}
and a new solution concept called \emph{intervention equilibrium}.
In an intervention game, the manager first chooses an intervention mechanism,
and then users choose their actions knowing the intervention mechanism chosen by the manager.
After observing a signal about the actions of users,
the manager chooses its action according to the intervention mechanism.
An intervention mechanism specifies an action that the manager takes following each
possible signal realization, and thus can be considered as a protocol that prescribes a rule
according to which the manager rewards and punishes users depending on their observed behavior.
The manager chooses an intervention mechanism to optimize its objective, anticipating
the rational behavior of users given the intervention mechanism it chooses.
Intervention equilibrium predicts the outcome of an intervention game in terms of
an intervention mechanism designed by the manager and an operating point chosen
by users.

\subsection{Related Work}

An approach similar to ours can be found in \cite{garg}. The authors of \cite{garg}
consider a congestion control problem and analyze performance under different scheduling mechanisms,
which can be considered as intervention mechanisms.
With a scheduling mechanism that assigns a higher priority to flows with a smaller rate, the input rates
of users can be restrained voluntarily. In other words, users do not send their traffic
excessively in their self-interest because doing so will result in a higher
probability that their packets are dropped. Another closely related work is
our previous work \cite{jpark}. In \cite{jpark}, we consider a random
access network where the manager can intervene by jamming the packets of users and
show that intervention mechanisms can successfully control
the transmission probabilities of selfish users. However, these works
consider only specific communication networks.
In this paper, we aim to develop
a general framework of intervention mechanisms that can be applied to various
application scenarios including not only communication networks but also
other scenarios such as file sharing on peer-to-peer networks,
multi-task processing on cloud computers, and load balancing on multi-core processors.

In an intervention game, the manager makes a decision before users do,
and thus the manager can be considered as a leader and users as followers.
Such a hierarchical structure of multi-user interaction has been modeled
as a Stackelberg game in the context of congestion control \cite{korilis}, medium access
control \cite{ma}, and power control \cite{yisu2}.
The main difference between our intervention approach and the traditional
Stackelberg approach can be found in the ability of the manager. In an
intervention game, the manager can monitor the actions of users and
can commit to an intervention mechanism. Hence, a strategy of the manager
is its complete contingent plan for actions to be taken given a signal
realization. On the other hand, in the Stackelberg games in \cite{korilis}--\cite{yisu2},
the leader does not have monitoring and
commitment capabilities, and thus simply chooses an action before followers
choose their actions. Our formulation includes the traditional
Stackelberg formulation as a special case, because an intervention mechanism
reduces to an action if the manager chooses the same action regardless of
signal realizations or if it observes no signal. Therefore, intervention games
yield higher design flexibility for the manager than traditional Stackelberg
games do, enabling the greater potential to shape the incentives of users.

Pricing mechanisms that charge users for their usage have received
a significant amount of attention in the literature \cite{varian}. Pricing
mechanisms have a solid theoretical foundation in economics as well
as high design flexibility. They can be considered as
a special class of intervention mechanisms, where the intervention of
the manager corresponds to charging and collecting payments from users.
The existing literature has studied pricing mechanisms using
Stackelberg games in the context of congestion control \cite{basar}
and cognitive radio networks \cite{bloem}.

\subsection{Focus and Advantages of Our Approach}

Our discussion so far suggests a classification of intervention
into direct and indirect intervention. Intervention
is direct (resp. indirect) if the manager interacts with users in the
inside (resp. outside) of the network. In other words, the manager
with direct intervention intervenes in the network usage of users, while
the manager with indirect intervention influences the utilities of users
through an outside instrument, for example, monetary payments in the
case of pricing. Direct intervention can be further classified into
adaptive and constant intervention, depending on whether the manager
can react to the actions of users. Traditional Stackelberg strategies
where the manager takes an action before users do belong to constant
intervention. Table~\ref{table:class} classifies the existing
intervention approaches in the literature according to the above criteria.

\begin{table}
\caption{Classification of the Existing Intervention Approaches in the Literature.}
\centering
\begin{tabular}{|>{\centering\arraybackslash}p{0.8in}|>{\centering\arraybackslash}p{0.8in}|>{\centering\arraybackslash}p{1.2in}|}  
\hline
\multicolumn{2}{|>{\centering\arraybackslash}p{1.6in}|}{Direct intervention} & Indirect intervention \\ \cline{1-2}
Adaptive & Constant & (e.g., pricing) \\ \hline
\cite{garg}, \cite{jpark} & \cite{korilis}, \cite{ma}, \cite{yisu2} & \cite{basar}, \cite{bloem} \\
\hline
\end{tabular}
\label{table:class}
\end{table}

Although our framework covers all the three classes of intervention in Table~\ref{table:class},
our main focus is on adaptive direct intervention, which has not received
much attention from researchers compared to the other two classes.
Adaptive direct intervention is useful compared to
constant direct intervention especially when the manager
does not value its usage of the network. In this scenario, the manager
desires to achieve a certain operating point with the minimum level
of intervention. With constant direct intervention the manager
needs to consume network resources in order to impact the behavior
of users, whereas with adaptive direct intervention the manager
can use intervention only as a threat to punish users in case of
deviation. Direct intervention has an advantage over indirect
intervention in terms of implementation. Since direct intervention
affects the network usage of users directly, users cannot evade intervention
as long as they use the network. On the contrary, enforcing the outside
instrument of indirect intervention can be costly especially when
the network is freely accessible. The difference between direct and
indirect intervention yields an informational advantage of direct intervention
over indirect intervention. For instance, direct intervention in \cite{garg}
and \cite{jpark} affects the data rates of users. Hence, the effectiveness
of a direct intervention mechanism is independent of the way users value their
data rate. On the contrary, designing a pricing mechanism to achieve
a certain operating point in general requires the manager to know
the utility functions of users. This point is illustrated with an example
in Section~\ref{sec:app}.

\subsection{Comparison with Other Frameworks}

We first compare the framework of intervention mechanism design with that of
network utility maximization (NUM) \cite{low}.
Since our framework allows a general form of the objective of the manager,
the objective can be set as the sum of the utilities of users, as in NUM. The main difference between
intervention and NUM is that intervention takes into account the incentives of selfish users whereas
NUM assumes obedient users. The NUM framework aims to design a distributed
algorithm following which obedient users can reach a system-wide optimal operating point, using prices as congestion
signals. On the contrary, the intervention framework aims to design an incentive mechanism that induces
selfish users to achieve an incentive-constrained optimal operating point.
Hence, the intervention framework is more relevant
in a network with selfish users.

Next, we compare intervention games and repeated games \cite{mailath},
which share a common goal of sustaining cooperation among selfish individuals. With a
repeated game strategy such as tit-for-tat,
users monitor the actions of other users and choose their actions depending on their past
observations. Hence, the burden of monitoring and executing reward and punishment
is distributed to users in a repeated game, while it is imposed solely on the manager
in an intervention game. If there is no reliable manager in the network, a mechanism
based on a repeated game strategy is a possible alternative to an intervention mechanism.
However, in order to implement a repeated game strategy, users in the network must
interact frequently and maintain histories of their observations. Moreover, sustaining
an optimal operating point with a repeated game strategy often requires users to be
sufficiently patient. On the contrary, implementing intervention mechanisms effectively
requires neither repeated interaction nor patience of users.

Lastly, we compare intervention mechanism design with standard mechanism design
\cite{mas}, \cite{nisan}. A standard mechanism design problem considers a scenario where
the manager can control the system configuration but has incomplete information
about the types of users. Thus, standard mechanism design is concerned
about incentives regarding types, whereas intervention mechanism design in
this paper cares about incentives regarding actions. In economics terminology,
standard mechanism design focuses on \emph{adverse selection} while our intervention
mechanism design focuses on \emph{moral hazard} \cite{mas}. We can modify or extend our
framework to study the capabilities and limitations of intervention mechanisms
to overcome incentive problems in the presence of only adverse selection
or both adverse selection and moral hazard. We leave these topics for future
research.

\subsection{Organization of the Paper}

The remainder of this paper is organized as follows. In Section II,
we formulate the framework of intervention mechanism design, defining
intervention games and intervention equilibrium. In Section III, we
classify the types of intervention and the manager to point out the
broad scope of our framework.
In Section IV, we consider a benevolent manager with perfect monitoring
and provide analytic results about intervention equilibrium.
In Section V, we apply our framework to random access networks
in order to illustrate the results. We conclude in Section VI.

\section{Framework}

We consider a multi-user network where users interact with each
other taking actions independently. There is a network
manager that can intervene in the interaction of users.
There are $N$ users in the network, and the set of users
is denoted by $\mathcal{N} = \{1,\ldots,N\}$. For convenience,
we call the manager user~$0$. To distinguish the users from the manager,
we sometimes refer to a user as a regular user. The regular users are indexed
from user $1$ to user $N$. The set of all the users including the manager
is denoted by $\mathcal{N}_0 = \mathcal{N} \cup \{0\}$.
Each user~$i \in \mathcal{N}_0$ chooses an action $a_i$ from the
set of actions available to it, which is denoted by $A_i$.
An action profile of the regular users is written as $a = (a_1, \ldots, a_N)$,
while the set of action profiles of the regular users is written
as $A = \prod_{i \in \mathcal{N}} A_i$. An action profile of the regular
users other than regular user~$i$ is written as $a_{-i} = (a_1,\ldots,
a_{i-1},a_{i+1}, \ldots,a_N)$ so that $a$ can be expressed as $a = (a_i,a_{-i})$.

The actions of the manager and the users jointly determine utilities they
receive. The utility function of user $i \in \mathcal{N}_0$ is denoted by
$u_i: A_0 \times A \rightarrow \mathbb{R}$.
That is, $u_i(a_0,a)$ represents the utility that
user $i$ receives when the manager chooses action $a_0$ and the users choose
an action profile $a$.
The manager is different from the regular users in that the manager is able
to monitor the actions of the regular users before it takes its action.
The monitoring ability of the manager is formally represented by a monitoring
technology $(S,\rho)$. $S$ is the set of all possible signals that the
manager can obtain, while $\rho$ is a mapping from $A$
to $\triangle(S)$, where $\triangle(S)$ is the set of probability distributions
over $S$. That is, $\rho(a)$ represents the probability distribution
of signals when the regular users choose $a$.

Since the manager takes its action after observing a signal,
a strategy for the manager is its complete contingent plan
for actions to be taken following all possible signal realizations
while a strategy for a regular user is simply its
choice of an action.\footnote{For expositional simplicity,
we restrict attention to pure strategies, although our formulation extends easily to the case
of randomized strategies.}
Thus, a strategy for the manager
can be represented by a function $f:S \rightarrow A_0$, which we
call an \emph{intervention mechanism}. That is, an intervention mechanism
specifies an action taken by the manager following each signal it can observe.
Let $\mathcal{F}$ be the set of
all intervention mechanisms, i.e., the set of all functions from $S$ to $A_0$.
We assume that the manager is able to commit to an intervention
mechanism and that the intervention mechanism chosen by the manager is known to the regular users when they choose their actions.\footnote{The information
about the intervention mechanism can be obtained by an agreed protocol specification
or by learning.}

To sum up, an \emph{intervention game} is summarized by the data
\begin{align*}
\mathcal{G} = \left\langle \mathcal{N}_0, (A_i)_{i \in \mathcal{N}_0}, (u_i)_{i \in \mathcal{N}_0},
(S,\rho) \right\rangle,
\end{align*}
and the timing of an intervention game can be described as follows.
\begin{enumerate}
\item The manager chooses an intervention mechanism $f \in \mathcal{F}$.
\item The regular users choose their actions $a \in A$ independently and simultaneously, knowing the intervention mechanism $f$
chosen by the manager.
\item The manager observes a signal $s \in S$, which is realized
following a probability distribution $\rho(a) \in \triangle(S)$.
\item The manager chooses its action $a_0 \in A_0$ according to $f$, i.e., $a_0 = f(s)$.
\item The users receive their utilities based on the chosen actions $(a_0,a) \in
A_0 \times A$.
\end{enumerate}

Consider an intervention mechanism $f \in \mathcal{F}$, and define a function $v_i^f: A \rightarrow \mathbb{R}$ by
\begin{align} \label{eq:vif}
v_i^f(a) = E_{\rho(a)} [u_i(f(s),a)],
\end{align}
for all $i \in \mathcal{N}$, where $E_{\rho(a)} [\cdot]$ denotes
expectation with respect to the random variable $s$ following the distribution
$\rho(a)$. Then $v_i^f(a)$ represents the expected utility
that regular user $i$ receives when the manager uses intervention mechanism
$f$ and the regular users choose action profile $a$.
An intervention mechanism $f$ induces a simultaneous game played by the regular
users, whose normal form representation is given by
\begin{align*}
\mathcal{G}_f = \left\langle \mathcal{N}, (A_i)_{i \in \mathcal{N}}, (v_i^f)_{i \in \mathcal{N}} \right\rangle.
\end{align*}
We can predict actions chosen by the selfish regular users
given an intervention mechanism $f$ by applying the solution concept of
Nash equilibrium to the induced game $\mathcal{G}_f$.

\begin{definition}
An intervention mechanism $f \in \mathcal{F}$ \emph{supports} the action
profile $a^* \in A$ of the regular users if $a^*$ is a Nash
equilibrium of the game $\mathcal{G}_f$, i.e.,
\begin{align*}
v_i^f(a_i^*,a_{-i}^*) \geq v_i^f(a_i,a_{-i}^*) \quad \text{for all $a_i \in A_i$, for all $i \in \mathcal{N}$}.
\end{align*}
Also, $f$ \emph{strongly supports} $a^*$ if $a^*$ is a
unique Nash equilibrium of $\mathcal{G}_f$.
\end{definition}

If an action profile $a^*$ is supported by an intervention mechanism $f$,
the regular users cannot gain by a unilateral deviation from $a^*$ as long as the manager
uses intervention mechanism $f$. When choosing an intervention mechanism, the manager
can expect that the regular users will choose an action profile that is supported
by the intervention mechanism it chooses. However, when there are multiple action
profiles supported by the intervention mechanism, the manager may be uncertain
about the action profile chosen by the regular users. This uncertainty
disappears when the intervention mechanism strongly supports an action profile.
In the formulation of our solution concept, we assume that the regular users will always select the best
Nash equilibrium for the manager if the induced game has multiple
Nash equilibria.

\begin{definition} \label{def:ie}
$(f^*,a^*) \in \mathcal{F} \times A$ is an
(strong) equilibrium of intervention game $\mathcal{G}$, or an \emph{(strong) intervention equilibrium},
if $f^*$ (strongly) supports $a^*$ and
\begin{align*}
E_{\rho(a^*)} [u_0(f^*(s),a^*)] \geq E_{\rho(a)} [u_0(f(s),a)]  \quad \text{for all $(f,a) \in \mathcal{F} \times A$ such that $f$ supports $a$}.
\end{align*}
$f^* \in \mathcal{F}$ is an \emph{(strongly) optimal intervention mechanism} if there exists an
action profile $a^* \in A$
such that $(f^*,a^*)$ is an (strong) intervention equilibrium.
\end{definition}

Intervention equilibrium is a solution concept for
intervention games, based on a backward induction argument, assuming that
the manager can commit to an intervention mechanism and predict the rational reaction
of the regular users to its choice of an intervention mechanism. An intervention
equilibrium can be considered as a Stackelberg equilibrium (or a subgame perfect
equilibrium) applied to an intervention game $\mathcal{G}$, since the induced
game $\mathcal{G}_f$ is a subgame of $\mathcal{G}$.
Since we assume that the manager can induce the regular users
to choose the best Nash equilibrium for it in the case of multiple
Nash equilibria, the problem of designing
an optimal intervention mechanism can be expressed as
$\max_{f \in \mathcal{F}}
\max_{a \in \mathcal{E}(f)} E_{\rho(a)} [u_0(f(s),a)]$,
where $\mathcal{E}(f)$ denotes the set of action profiles supported by $f$.

\section{Classification of Intervention and the Manager}

In this section, we classify the types of intervention depending on
the form of the utility functions of the regular users (i.e., the way
the manager interacts with the regular users) and the types
of the manager depending on the form of the utility function the manager
(i.e., the objective of the manager). The purpose of classification is
to suggest that a wide range of application scenarios can be modeled
in the framework of intervention.

\begin{definition}
Let $A_i \subset \mathbb{R}^K$ for some $K$, for all $i \in \mathcal{N}_0$.
Intervention is \emph{additively symmetric} if, for each $i \in \mathcal{N}$, there exist functions
$g_{ij}^k:\mathbb{R} \rightarrow \mathbb{R}$ and
$h_i^k:\mathbb{R}^2 \rightarrow \mathbb{R}$, for $k =1,\ldots,K$ and for $j \in \mathcal{N}_0 \setminus \{i\}$,
such that
\begin{align*}
u_i(a_0,a) = \sum_{k=1}^K h_i^k\left(a_i^k, \sum_{j \in \mathcal{N}_0 \setminus \{i\}} g_{ij}^k(a_j^k) \right),
\end{align*}
where $a_i^k$ denotes the $k$-th element of $a_i$.
Intervention is \emph{multiplicatively symmetric} if, for each $i \in \mathcal{N}$, there exist functions
$g_{ij}^k:\mathbb{R} \rightarrow \mathbb{R}$ and
$h_i^k:\mathbb{R}^2 \rightarrow \mathbb{R}$, for $k =1,\ldots,K$ and for $j \in \mathcal{N}_0 \setminus \{i\}$,
such that
\begin{align*}
u_i(a_0,a) = \sum_{k=1}^K h_i^k\left(a_i^k, \prod_{j \in \mathcal{N}_0 \setminus \{i\}} g_{ij}^k(a_j^k) \right).
\end{align*}
Intervention is \emph{symmetric} if it is either additively or multiplicatively symmetric.
\end{definition}

The manager with symmetric intervention uses the network as the regular users do,
in the same position as the regular users.
The effect of the actions of other uses including the manager on a regular user is represented by an
additive term in the case of additive symmetric intervention and a multiplicative
term in the case of multiplicatively symmetric intervention.
Network models in which the manager can exert additive symmetric intervention
include routing \cite{korilis} and frequency-selective Gaussian interference channels \cite{yisu2}.
Multiplicative symmetric intervention can be used in the random access model of \cite{jpark}.

We can interpret the objective of the manager as the system objective, which
can vary depending on the types of the manager.
Below we classify the manager with symmetric intervention depending on the form of its utility function.

\begin{definition}
Consider symmetric intervention.
The manager is \emph{benevolent} if
\begin{align*}
u_0(a_0,a) = \sum_{i \in \mathcal{N}} w_i u_i(a_0,a)
\end{align*}
for some $(w_1, \ldots, w_N) \in \mathbb{R}_+^N$.
The manager is \emph{self-interested} if there exist functions
$g_{0j}^k:\mathbb{R} \rightarrow \mathbb{R}$ and
$h_0^k:\mathbb{R}^2 \rightarrow \mathbb{R}$, for $k =1,\ldots,K$ and for $j \in \mathcal{N}$,
such that
\begin{align} \label{eq:selfadd}
u_0(a_0,a) = \sum_{k=1}^K h_0^k\left(a_i^k, \sum_{j \in \mathcal{N}} g_{0j}^k(a_j^k) \right)
\end{align}
in the case of additively symmetric intervention, and
\begin{align} \label{eq:selfmulti}
u_0(a_0,a) = \sum_{k=1}^K h_0^k\left(a_i^k, \prod_{j \in \mathcal{N}} g_{0j}^k(a_j^k) \right)
\end{align}
in the case of multiplicatively symmetric intervention.
The manager is \emph{total welfare maximizing} if
\begin{align*}
u_0(a_0,a) = w_0 \tilde{u}_0(a_0,a) + \sum_{i \in \mathcal{N}} w_i u_i(a_0,a)
\end{align*}
for some $(w_0, w_1, \ldots, w_N) \in \mathbb{R}_+^{N+1}$, where $\tilde{u}_0(a_0,a)$
denotes the right-hand side of \eqref{eq:selfadd} if intervention is additively symmetric
and that of \eqref{eq:selfmulti} if intervention is multiplicatively symmetric.
\end{definition}

A benevolent manager maximizes the welfare of the regular users, attaching a welfare
weight $w_i$ to regular user $i$.
The utility function of a self-interested manager has the same
structure as those of the regular users.
Recall that the manager with symmetric intervention uses the network
as the regular users do. A benevolent manager does not derive any
utility from its usage of the network, whereas the utility of a self-interested
manager is derived solely from its usage \cite{yisu2}.
A total welfare maximizing manager derives utility from its usage
as well as that of the other users \cite{korilis}.

It is also possible that the manager plays a special role in the network,
participating in interaction in a different way from the regular users.
For example, the manager plays the role of a scheduler in \cite{garg}
and a billing authority in pricing \cite{basar}, \cite{bloem}.
We call this type of intervention asymmetric intervention to contrast
it with symmetric intervention.\footnote{Indirect intervention is necessarily
asymmetric, while direct intervention can be symmetric or asymmetric.}
We present two representative examples of asymmetric intervention,
which arise naturally in games with transferable utility \cite{myerson}.

\begin{definition} \label{def:asymmetric}
Intervention is \emph{asymmetric} if it is not symmetric.
Let $A_0 = \mathbb{R}_+^N$.
Intervention is \emph{additively asymmetric} if, for each $i \in \mathcal{N}$, there exists function
$g_{i}:A \rightarrow \mathbb{R}$ such that
\begin{align*}
u_i(a_0,a) = g_i(a) - a_0^i,
\end{align*}
where $a_0^i$ is the $i$-th element of $a_0$.
Let $A_0 = [0,1]^N$.
Intervention is \emph{multiplicatively asymmetric} if, for each $i \in \mathcal{N}$, there exists function
$g_{i}:A \rightarrow \mathbb{R}_+$ such that
\begin{align*}
u_i(a_0,a) = (1-a_0^i) g_i(a).
\end{align*}
\end{definition}

With additively asymmetric intervention, the manager takes away the amount $a_0^i$
from the benefit $g_i(a)$ that regular user $i$ receives from its usage of the network.
With multiplicatively asymmetric intervention, the manager deducts
the $a_0^i$ fraction of the benefit that regular user $i$ receives.
Hence, the two types of asymmetric intervention in Definition~\ref{def:asymmetric} can be compared
to two different forms of taxation, where additively asymmetric intervention
corresponds to taxation with lump-sum tax rates and multiplicatively asymmetric intervention
to taxation with proportional tax rates.

\begin{definition} \label{def:asymman}
Consider additively or multiplicatively asymmetric intervention. Let $c_0:A \rightarrow \mathbb{R}_+$
be a function that represents the operating cost of the network.
The manager is \emph{benevolent} if
\begin{align*}
u_0(a_0,a) = \sum_{i \in \mathcal{N}} u_i(a_0,a).
\end{align*}
The manager is \emph{self-interested} if
\begin{align*}
u_0(a_0,a) = \sum_{i \in \mathcal{N}} a_0^i - c_0(a)
\end{align*}
in the case of additively asymmetric intervention, and
\begin{align*}
u_0(a_0,a) = \sum_{i \in \mathcal{N}} a_0^i g_i(a) - c_0(a)
\end{align*}
in the case of multiplicatively asymmetric intervention.
The manager is \emph{total welfare maximizing} if
\begin{align*}
u_0(a_0,a) = \sum_{i \in \mathcal{N}} g_i(a) - c_0(a).
\end{align*}
\end{definition}

In Definition~\ref{def:asymman}, we use welfare weights $w_i = 1$ for all $i \in \mathcal{N}_0$
because of the assumption of transferable utility.
For ease of interpretation, let us regard the transfer from the regular users to the manager
as payments.
A benevolent manager does not value payments (or burns payments) it receives from the regular users,
and thus payments create a welfare loss.
On the contrary, a self-interested manager values payments from the regular users,
and it maximizes its profit measured by the total payment minus the operating cost. If there is
no operating cost, i.e., $c_0 \equiv 0$, then the objective of a self-interested manager
is revenue maximization.
A total welfare maximizing manager maximizes the net gain from operating the network,
which is the total benefit of the regular users minus the operating cost.
We can also consider an \emph{individually rational benevolent} manager that has utility function
$u_0(a_0,a) = \sum_{i \in \mathcal{N}} u_i(a_0,a)$ and faces the
individual rationality constraint that requires the total payment it receives to be no less
than the operating cost (i.e., $\sum_{i \in \mathcal{N}} a_0^i \geq c_0(a)$
in the case of additively asymmetric intervention and $\sum_{i \in \mathcal{N}} a_0^i g_i(a)
\geq c_0(a)$ in the case of multiplicatively asymmetric intervention). In case that a regular
user has an outside option, the individual rationality constraint for the regular user can be
taken care of by including in its action space an action that corresponds to choosing its outside
option.

\section{Benevolent Manager with Perfect Monitoring} \label{sec:permon}

In this section, we analyze a class of intervention games that satisfy
the following maintained assumptions, while leaving the analysis of other
classes for future work.

\begin{assumption} \label{ass:permon}
(i) (Benevolent manager) The utility function of the manager is given by
$u_0(a_0,a) = \sum_{i \in \mathcal{N}} u_i(a_0,a)$ for all $(a_0,a) \in A_0 \times A$.\\
(ii) (Existence of minimal and maximal intervention actions) There exist the minimal and maximal
elements of $A_0$, denoted $\underline{a}_0$ and $\overline{a}_0$, respectively, in the sense that
for all $i \in \mathcal{N}$, $\underline{a}_0$ and $\overline{a}_0$ satisfy
\begin{align*}
u_i(\underline{a}_0,a) \geq u_i(a_0,a) \geq u_i(\overline{a}_0,a) \quad \text{for all $a_0 \in A_0$, for all $a \in A$}.
\end{align*}
(iii) (Perfect monitoring) The monitoring technology of the manager is
perfect in the sense that the manager can observe the actions of the regular users
without errors. Formally, a monitoring technology $(S, \rho)$ is perfect
if $S = A$ and only signal $a$ can arise in the distribution $\rho(a)$ for all $a \in A$.
\end{assumption}

In Assumption~\ref{ass:permon}(i), we set welfare weights as $w_i = 1$ for
all $i \in \mathcal{N}$ in order to cover both cases of transferable and nontransferable
utility, although our results extend easily to general welfare weights.
In Assumption~\ref{ass:permon}(ii), $\underline{a}_0$ and $\overline{a}_0$
can be interpreted as the minimal and maximal intervention actions of the manager,
respectively. For given $a \in A$, each regular user receives the highest (resp. lowest) utility
when the manager takes the minimal (resp. maximal) intervention action.
In other words, the utilities of the regular users are aligned with
respect to the action of the manager so that the manager can reward
or punish all the regular users at the same time.
Combining Assumption~\ref{ass:permon}(i) and (ii), we obtain
\begin{align} \label{eq:u0order}
u_0(\underline{a}_0,a) \geq u_0(a_0,a) \geq u_0(\overline{a}_0,a) \quad \text{for all $a_0 \in A_0$, for all $a \in A$}.
\end{align}
Thus, for given $a \in A$ the benevolent manager prefers to use the
minimal intervention action.
With perfect monitoring, $v_i^f$ defined in \eqref{eq:vif} reduces to
$v_i^f(a) = u_i(f(a),a)$ for all $a \in A$.

We first characterize the set of action profiles of the regular users
that can be supported by an intervention mechanism. Define
$\mathcal{E} = \cup_{f \in \mathcal{F}} \mathcal{E}(f) = \{ a \in A : \exists f \in \mathcal{F} \text{ such that $f$ supports $a$} \}$.
\begin{proposition} \label{prop:char}
$a^* \in \mathcal{E}$ if and only if $u_i(\underline{a}_0,a^*) \geq u_i(\overline{a}_0, a_i, a_{-i}^*)$ for all $a_i \in A_i$, for all $i \in \mathcal{N}$.
\end{proposition}

\begin{IEEEproof}
Suppose that $u_i(\underline{a}_0,a^*) \geq u_i(\overline{a}_0, a_i, a_{-i}^*)$ for all $a_i \in A_i$, for all $i \in \mathcal{N}$.
Define an intervention mechanism $f_{\tilde{a}}$, for each $\tilde{a} \in A$, by
\begin{align} \label{eq:polar}
f_{\tilde{a}}(a) = \left\{
\begin{array}{ll}
\underline{a}_0 \quad &\textrm{if $a = \tilde{a}$,}\\
\overline{a}_0 \quad &\textrm{otherwise}.
\end{array} \right.
\end{align}
Then $f_{a^*}$ supports $a^*$, and thus $a^* \in \mathcal{E}$.

Suppose that $a^* \in \mathcal{E}$. Then there
exists an intervention mechanism $f$ such that $u_i(f(a^*), a^*)
\geq u_i(f(a_i, a_{-i}^*), a_i, a_{-i}^*)$ for all $a_i \in A_i$, for all $i \in \mathcal{N}$.
Then we obtain $u_i(\underline{a}_0,a^*) \geq u_i(f(a^*), a^*)
\geq u_i(f(a_i, a_{-i}^*), a_i, a_{-i}^*) \geq u_i(\overline{a}_0, a_i, a_{-i}^*)$ for all $a_i \in A_i$, for all $i \in \mathcal{N}$,
where the first and the third inequalities follow from Assumption~\ref{ass:permon}(ii).
\end{IEEEproof}

The basic idea underlying Proposition~\ref{prop:char} is that for
given $a^* \in A$, $f_{a^*}$ is the most effective intervention mechanism
to support $a^*$. Thus, in order to find out whether $a^*$ is supported by
some intervention function, it suffices to check whether $a^*$ is supported
by $f_{a^*}$.
We call $f_{\tilde{a}}$, defined in \eqref{eq:polar}, the \emph{maximum punishment} intervention mechanism
with target action profile $\tilde{a}$, because the manager using $f_{\tilde{a}}$
takes the maximum intervention action whenever the users do not follow the action profile $\tilde{a}$.
Let $\mathcal{F}^p$ be the set of all maximum punishment intervention mechanisms, i.e.,
$\mathcal{F}^p = \{ f_{\tilde{a}} \in \mathcal{F} : \tilde{a} \in A \}$.
Also, define $\mathcal{E}^p = \cup_{f \in \mathcal{F}^p} \mathcal{E}(f) = \{ a \in A : \exists f \in \mathcal{F}^p \text{ such that $f$ supports $a$} \}$.
The second part of the proof of Proposition~\ref{prop:char} shows that
if $a^*$ is supported by some intervention mechanism $f$, then
it is also supported by the maximum punishment intervention mechanism with
target action profile $a^*$, $f_{a^*}$. This observation leads us to the following corollary.

\begin{corollary} \label{cor:polar}
(i) $\mathcal{E} = \mathcal{E}^p$.\\
(ii) If $(f^*,a^*)$ is an intervention equilibrium, then $(f_{a^*}, a^*)$ is also an intervention
equilibrium.
\end{corollary}

\begin{IEEEproof}
(i) $\mathcal{E} \supset \mathcal{E}^p$ follows from $\mathcal{F} \supset \mathcal{F}^p$,
while $\mathcal{E} \subset \mathcal{E}^p$ follows from Proposition~\ref{prop:char}.

(ii) Suppose that $(f^*,a^*)$ is an intervention equilibrium. Then by Definition~\ref{def:ie},
$f^*$ supports $a^*$, and $u_0(f^*(a^*),a^*) \geq u_0(f(a),a)$
for all $(f,a) \in \mathcal{F} \times A$ such that $f$ supports $a$.
Since $a^* \in \mathcal{E}$, $f_{a^*}$ supports $a^*$ by Proposition~\ref{prop:char}.
Hence, $u_0(f^*(a^*),a^*) \geq u_0(f_{a^*}(a^*),a^*)$. On the other hand,
since $f_{a^*}(a^*) = \underline{a}_0$, we have $u_0(f^*(a^*),a^*) \leq u_0(f_{a^*}(a^*),a^*)$ by \eqref{eq:u0order}.
Therefore, $u_0(f^*(a^*),a^*) = u_0(f_{a^*}(a^*),a^*)$, and thus
$u_0(f_{a^*}(a^*),a^*) \geq u_0(f(a),a)$
for all $(f,a) \in \mathcal{F} \times A$ such that $f$ supports $a$.
This proves that $(f_{a^*}, a^*)$ is an intervention
equilibrium.
\end{IEEEproof}

Corollary~\ref{cor:polar} shows that there is no loss of generality in two senses
when we restrict attention to maximum punishment intervention mechanisms.
First, the set of action profiles that can be supported
by an intervention mechanism remains the same when we consider
only maximum punishment intervention mechanisms. Second, if there exists
an optimal intervention mechanism, we can find an optimal intervention mechanism
among maximum punishment intervention mechanisms. The following proposition
provides a necessary and sufficient condition under which a maximum punishment intervention mechanism
together with its target action profile constitutes an intervention
equilibrium.

\begin{proposition} \label{prop:charie}
$(f_{a^*}, a^*)$ is an intervention equilibrium if and only if
$a^* \in \mathcal{E}$ and $u_0(\underline{a}_0, a^*) \geq
u_0(\underline{a}_0, a)$ for all $a \in \mathcal{E}$.
\end{proposition}

\begin{IEEEproof}
Suppose that $(f_{a^*}, a^*)$ is an intervention equilibrium.
Then $f_{a^*}$ supports $a^*$, and thus $a^* \in \mathcal{E}$.
Also, $u_0(f_{a^*}(a^*),a^*) \geq u_0(f(a),a)$
for all $(f,a) \in \mathcal{F} \times A$ such that $f$ supports $a$.
Choose any $a \in \mathcal{E}$. Then by Proposition~\ref{prop:char},
$f_a$ supports $a$, and thus $u_0(\underline{a}_0, a^*) = u_0(f_{a^*}(a^*),a^*)
\geq u_0(f_a(a),a) = u_0(\underline{a}_0, a)$.

Suppose that $a^* \in \mathcal{E}$ and $u_0(\underline{a}_0, a^*) \geq
u_0(\underline{a}_0, a)$ for all $a \in \mathcal{E}$.
To prove that $(f_{a^*},a^*)$ is an intervention equilibrium, we need to
show (i) $f_{a^*}$ supports $a^*$, and (ii) $u_0(f_{a^*}(a^*),a^*) \geq u_0(f(a),a)$
for all $(f,a) \in \mathcal{F} \times A$ such that $f$ supports $a$.
Since $a^* \in \mathcal{E}$, (i) follows from Proposition~\ref{prop:char}.
To prove (ii), choose any $(f,a) \in \mathcal{F} \times A$ such that $f$ supports $a$.
Then $u_0(f_{a^*}(a^*),a^*) = u_0(\underline{a}_0, a^*) \geq u_0(\underline{a}_0, a) \geq u_0(f(a),a)$,
where the first inequality follows from $a \in \mathcal{E}$.
\end{IEEEproof}

Proposition~\ref{prop:charie} implies that if we obtain an action profile
$a^*$ such that $a^* \in \arg \max_{a \in \mathcal{E}} u_0(\underline{a}_0, a)$,
we can use it to construct an intervention equilibrium and thus an optimal
intervention mechanism. Corollary~\ref{cor:polar}(ii) implies that,
when we want to find out whether a given action profile can be supported
by an optimal intervention mechanism, we can consider only the maximum punishment
intervention mechanism having the action profile as its target action profile. However,
when we are given an optimal intervention mechanism $f_{a^*}$ in the class of
maximum punishment intervention mechanisms,
it is not certain whether its target action profile $a^*$ or some
other action profile constitutes an intervention equilibrium together with $f_{a^*}$.
The following proposition provides a sufficient condition under
which a given optimal intervention mechanism $f_{a^*}$
must be paired with its target action profile $a^*$ to form an intervention
equilibrium.

\begin{proposition} \label{prop:optie}
Suppose that $u_0(\underline{a}_0, a) >
u_0(\overline{a}_0, a)$ for all $a \in \mathcal{E}$.
If $f_{a^*} \in \mathcal{F}^p$ is an optimal intervention mechanism,
then $(f_{a^*},a^*)$ is an intervention equilibrium and there exists no other $a \neq a^*$
such that $(f_{a^*},a)$ is an intervention equilibrium.
\end{proposition}

\begin{IEEEproof}
Suppose that $f_{a^*} \in \mathcal{F}^p$ is an optimal intervention mechanism.
Then there must exist $a' \in A$ such that $(f_{a^*}, a')$ is an intervention
equilibrium, i.e., (i) $f_{a^*}$ supports $a'$, and
(ii) $u_0(f_{a^*}(a'),a') \geq u_0(f(a),a)$
for all $(f,a) \in \mathcal{F} \times A$ such that $f$ supports $a$.
Suppose that there exists $a' \neq a^*$ that satisfies (i) and (ii).
Since $f_{a^*}$ supports $a'$, we have $a' \in \mathcal{E}$ and thus $f_{a'}$
supports $a'$. Then, by (ii), $u_0(\overline{a}_0,a') = u_0(f_{a^*}(a'),a') \geq
u_0(f_{a'}(a'),a') = u_0(\underline{a}_0,a')$, which contradicts the assumption
that $u_0(\underline{a}_0, a) > u_0(\overline{a}_0, a)$ for all $a \in \mathcal{E}$.
Therefore, there cannot exist $a' \neq a^*$ that satisfies (i) and (ii).
Since there must exist $a' \in A$ that satisfies (i) and (ii),
$a^*$ must satisfy (i) and (ii).
\end{IEEEproof}

Although maximum punishment intervention mechanisms have a simple structure and
are most effective in supporting a given action profile, they also have
weaknesses. First, even when $(f_{a^*},a^*)$ is an intervention equilibrium,
there may be other action profiles that are also supported by $f_{a^*}$.
For example, in the case of multiplicatively asymmetric intervention
where we have $\underline{a}_0 = (0,\ldots,0)$ and $\overline{a}_0 = (1,\ldots,1)$,
$f_{a^*}$ supports any action profile $a'$ that has at least two different
elements from those of $a^*$. At such an action profile, the regular users
receive zero utility. As long as $u_0(\underline{a}_0,a^*) > 0$, $a^*$ Pareto dominates other action
profiles that are supported by $f_{a^*}$, and this can provide a rationale that the regular
users will select $a^*$ among multiple Nash equilibria of the game induced by $f_{a^*}$.
However, in principle, the regular users may select any
Nash equilibrium, and the manager cannot
guarantee that the regular users will choose the intended target action profile $a^*$
when there are other Nash equilibria. Strongly optimal intervention mechanisms have
robustness in that they yield a unique Nash equilibrium and thus the issue of
multiple Nash equilibria does not arise.
As mentioned in Section II, an implicit assumption underlying the concept of intervention equilibrium
is that the regular
users will always select the best Nash equilibrium for the manager in the induced game.
If the manager takes a conservative approach, it may assume that the regular
users will choose the worst Nash equilibrium for it. In this
case, an optimal intervention mechanism solves $\max_{f \in \mathcal{F}}
\min_{a \in \mathcal{E}(f)} E_{\rho(a)} [u_0(f(s),a)]$,
which has a similar spirit as maximin strategies \cite{myerson} in non-zero-sum games.

Another weakness of maximum punishment intervention mechanisms is that they may
incur a large efficiency loss when there are errors in the system. For example,
when a regular user chooses an action different from the one it intends to take
by mistake (i.e., trembling hand) or when the manager receives an incorrect signal (i.e., noisy observation),
the maximal intervention action is applied even if the regular users (intend to)
choose the target action profile.
The case of noisy observation can be covered by modeling the monitoring technology of the manager as imperfect monitoring.
In order to overcome this weakness in the case of perfect monitoring,
we can consider a class of continuous intervention mechanisms (i.e., intervention
mechanisms represented by a continuous function from $A$ to $A_0$) if
the action spaces are continua.\footnote{Following \cite{mailath}, we say that
an action space is a continuum if it is a compact and convex subset of the Euclidean
space $\mathbb{R}^K$ for some $K$.} To obtain a concrete result,
we consider an intervention game where $A_i = [\underline{a}_i, \overline{a}_i] \subset
\mathbb{R}$ with $\underline{a}_i < \overline{a}_i$ for all $i \in \mathcal{N}_0$ and
$u_i(a_0,a)$ is strictly decreasing in $a_0$ on $[\underline{a}_0, \overline{a}_0]$
for all $a \in A$, for all $i \in \mathcal{N}$.
We define an intervention mechanism $f_{\tilde{a},c}$, for each $\tilde{a} \in A$ and $c \in \mathbb{R}^N$, by
\begin{align*}
f_{\tilde{a},c}(a) = \left[ c \cdot (a - \tilde{a}) \right]_{\underline{a}_0}^{\overline{a}_0},
\end{align*}
for all $a \in A$, where $[x]_{\alpha}^{\beta} = \min \{ \max\{x,\alpha\}, \beta\}$.
We call $f_{\tilde{a},c}$ the \emph{affine} intervention mechanism with
target action profile $\tilde{a}$ and intervention rate profile $c$.
The following proposition constructs an affine intervention mechanism
to support an interior target action profile in the case of differentiable utility functions.

\begin{proposition} \label{prop:affine}
Suppose that $u_i$ is twice continuously differentiable for all $i \in \mathcal{N}$.
Let $a^* \in A$ be an action profile such that $a_i^* \in (\underline{a}_i, \overline{a}_i)$ for all $i \in \mathcal{N}$,
and let
\begin{align} \label{eq:ci}
c_i^* = - \frac{\partial u_i(\underline{a}_0, a^*)/\partial a_i}{\partial u_i(\underline{a}_0, a^*)/\partial a_0}
\end{align}
for all $i \in \mathcal{N}$.\footnote{We define $\partial u_i(\underline{a}_0, a^*)/\partial a_0$ as the right partial derivative of
$u_i$ with respect to $a_0$ at $(\underline{a}_0, a^*)$.} Suppose that
\begin{align} \label{eq:soc1}
\frac{\partial^2 u_i}{\partial a_i^2} (\underline{a}_0, a_i, a_{-i}^*) \leq 0 \quad \text{for all $a_i \in (\underline{a}_i, \overline{a}_i)$}
\end{align}
for all $i \in \mathcal{N}$ such that $c_i^* = 0$,
\begin{align} \label{eq:soc21}
\frac{\partial^2 u_i}{\partial a_i^2} (\underline{a}_0, a_i, a_{-i}^*) \leq 0 \quad \text{for all $a_i \in (\underline{a}_i, a_i^*)$},
\end{align}
\begin{align} \label{eq:soc22}
\left( (c_i^*)^2 \frac{\partial^2 u_i}{\partial a_0^2}
+ 2 c_i^* \frac{\partial^2 u_i}{\partial a_i \partial a_0}
+ \frac{\partial^2 u_i}{\partial a_i^2} \right) \bigg|_{(a_0, a_i, a_{-i}) = (c_i^* (a_i - a_i^*) + \underline{a}_0, a_i, a_{-i}^*)} \leq 0 \nonumber \\
\quad \text{for all $a_i \in (a_i^*, \min \{\overline{a}_i, a_i^* + (\overline{a}_0 - \underline{a}_0)/c_i^* \})$},
\end{align}
\begin{align} \label{eq:soc23}
\frac{\partial u_i}{\partial a_i} (\overline{a}_0, a_i, a_{-i}^*) \leq 0 \quad \text{for all $a_i
\in (a_i^* + (\overline{a}_0 - \underline{a}_0)/c_i^*, \overline{a}_i)$}
\end{align}
for all $i \in \mathcal{N}$ such that $c_i^* > 0$, and
\begin{align*} 
\frac{\partial u_i}{\partial a_i} (\overline{a}_0, a_i, a_{-i}^*) \geq 0 \quad \text{for all $a_i \in
(\underline{a}_i, a_i^* + (\overline{a}_0 - \underline{a}_0)/c_i^*)$},
\end{align*}
\begin{align} \label{eq:soc32}
\left( (c_i^*)^2 \frac{\partial^2 u_i}{\partial a_0^2}
+ 2 c_i^* \frac{\partial^2 u_i}{\partial a_i \partial a_0}
+ \frac{\partial^2 u_i}{\partial a_i^2} \right) \bigg|_{(a_0, a_i, a_{-i}) = (c_i^* (a_i - a_i^*) + \underline{a}_0, a_i, a_{-i}^*)} \leq 0 \nonumber \\
\quad \text{for all $a_i \in (\max \{\overline{a}_i, a_i^* + (\overline{a}_0 - \underline{a}_0)/c_i^* \}, a_i^*)$},
\end{align}
\begin{align*} 
\frac{\partial^2 u_i}{\partial a_i^2} (\underline{a}_0, a_i, a_{-i}^*) \leq 0 \quad \text{for all $a_i
\in (a_i^*, \overline{a}_i)$}
\end{align*}
for all $i \in \mathcal{N}$ such that $c_i^* < 0$.\footnote{We define $(\alpha, \beta) = \emptyset$ if $\alpha \geq \beta$.}
Then $f_{a^*,c^*}$ supports $a^*$.
\end{proposition}

\begin{IEEEproof}
See the Appendix.
\end{IEEEproof}

Note that $\partial u_i(\underline{a}_0, a^*)/\partial a_0 < 0$ for all $i \in \mathcal{N}$
since $u_i$ is strictly decreasing in $a_0$. Thus, $c_i^*$, defined in \eqref{eq:ci}, has
the same sign as $\partial u_i(\underline{a}_0, a^*)/\partial a_i$. With $A_0 = [\underline{a}_0,
\overline{a}_0]$, the action of the manager can be interpreted
as the intervention level, and the regular users receive higher utility as the intervention
level is smaller.
The affine intervention mechanism $f_{a^*,c^*}$, constructed in Proposition~\ref{prop:affine},
has the properties that the manager chooses the minimal intervention level $\underline{a}_0$
when the regular users choose the target action profile $a^*$, i.e., $f_{a^*,c^*}(a^*) = \underline{a}_0$,
and that the intervention level increases in the rate of $|c_i^*|$ as regular
user $i$ deviates to the direction in which its utility increases at $(\underline{a}_0, a^*)$.
The expression of $c_i^*$ in \eqref{eq:ci} has an intuitive explanation.
Since $c_i^*$ is proportional to $\partial u_i(\underline{a}_0, a^*)/\partial a_i$
and inversely proportional to $-\partial u_i(\underline{a}_0, a^*)/\partial a_0$,
a regular user faces a higher intervention rate as its incentive to deviate
from $(\underline{a}_0, a^*)$ is stronger and as a change in the intervention level has a smaller
impact on its utility. The intervention level does not react to the action of regular user $i$
when $c_i^* = 0$, because regular user $i$ chooses $a_i^*$ in its self-interest
even when the intervention level is fixed at $\underline{a}_0$, provided that other regular
users choose $a_{-i}^*$. Finally, we note that if $(f_{a^*},a^*)$ is an
intervention equilibrium and $f_{a^*,c}$ supports $a^*$ for some $c$,
then $(f_{a^*,c},a^*)$ is also an intervention equilibrium, since $f_{a^*}(a^*) =
f_{a^*,c}(a^*) = \underline{a}_0$.

\section{Application to Random Access Networks} \label{sec:app}

In this section, we illustrate the results of Section~\ref{sec:permon}
by introducing a manager in a model of random access networks,
similar to the model of \cite{hamed}.
Time is divided into slots of equal length, and a user can transmit
its packet or wait in each slot. Due to interference, a packet
is successfully transmitted only if there is no other packet
transmitted in the current slot. If more than one packet
is transmitted simultaneously, a collision occurs.
The manager can transmit its packets as users do, and it interferes with all the users.

We model the random access scenario as an intervention game.
We assume that each user, including the manager, transmits its packets with a fixed
probability over time. The action of user $i$, $a_i$, is thus
its transmission probability, and we have $A_i = [0,1]$ for all $i \in \mathcal{N}_0$.
The average data rate for user $i \in \mathcal{N}$ when the users transmit according
to the probabilities $(a_0,a) \in A_0 \times A$ is given by
\begin{align*}
r_i(a_0,a) = \gamma_i a_i \prod_{j \in \mathcal{N}_0 \setminus \{i\}} (1-a_j),
\end{align*}
where $\gamma_i > 0$ is the fixed peak data rate for user $i$.
The benefit that a regular user $i$ obtains from its average data rate
is represented by a utility function $U_i : \mathbb{R}_+ \rightarrow \mathbb{R}$,
which is assumed to be strictly increasing.
Hence, the utility function of regular user $i$ in the intervention game is
given by $u_i(a_0,a) = U_i(r_i(a_0,a))$. We assume that the manager is benevolent
with the utility function $u_0(a_0,a) = \sum_{i \in \mathcal{N}} u_i(a_0,a)$
and that its monitoring technology is perfect.

In the above intervention game, the minimal and maximal intervention actions are given by
$\underline{a}_0 = 0$ and $\overline{a}_0 = 1$, respectively.
Since $r_i(0,a) \geq 0$ and $r_i(1,a) = 0$ for all $a \in A$, for all
$i \in \mathcal{N}$, we have $\mathcal{E} = A$ by Proposition~\ref{prop:char}.
In other words, any action profile $a \in A$ can be supported by an
intervention mechanism.
Because the maximal intervention action yields
zero rate to all the users regardless of the action profile, the maximum punishment is strong
enough to prevent deviations from any target action profile.\footnote{In fact, a maximum punishment
intervention mechanism can prevent not only unilateral deviations from its target
action profile but also joint deviations.}
Since $A=[0,1]^N$ is compact and $u_0$ is continuous,
a solution to $\max_{a \in A} u_0(0,a)$ exists.
Then by Proposition~\ref{prop:charie},
$(f_{a^*},a^*)$ is an intervention equilibrium if and only if $a^*$ maximizes
$u_0(0,a)$ on $A$. Also, we can apply Proposition~\ref{prop:affine}
to show that any $a^* \in (0,1)^N$ is supported by
$f_{a^*,c^*}$ with $c_i^* = 1/a_i^*$ for all $i \in \mathcal{N}$.
Suppose for the moment that the utility of each user is given by
its average data rate, i.e., $u_i(a_0,a) = r_i(a_0,a)$ for all $i \in \mathcal{N}$.
Then for each $i \in \mathcal{N}$, we obtain $c_i^* = 1/a_i^* > 0$ by
\eqref{eq:ci}, and we can verify that the conditions \eqref{eq:soc21}--\eqref{eq:soc23}
are satisfied. Thus, by Proposition~\ref{prop:affine}, we can conclude that $f_{a^*,c^*}$
supports $a^*$. Note that the concept of an intervention mechanism supporting an
action profile is based on Nash equilibrium, which uses only the ordinal
properties of the utility functions. Therefore, $f_{a^*,c^*}$ still supports $a^*$
even when the utility function of user $i$ is given by $u_i(a_0,a) = U_i(r_i(a_0,a))$
for any strictly increasing function $U_i$, for all $i \in \mathcal{N}$.

The above argument points out an informational advantage of direct intervention
over indirect intervention. To highlight the informational advantage of direct intervention, suppose that
the objective of the manager is to implement a target action profile $a^* \in (0,1)^N$, determined
independently of $(U_1,\ldots,U_N)$, while taking the minimal intervention action
when the users choose $a^*$. Then the results in the previous paragraph imply
that the direct intervention mechanisms $f_{a^*}$ and $f_{a^*,c^*}$ with
$c_i^* = 1/a_i^*$ for all $i \in \mathcal{N}$ support the action profile $a^*$
for \emph{any} $(U_1,\ldots,U_N)$. Since direct intervention affects utility
through its impact on the rates, its effectiveness does not depend on the shapes of the utility
functions.
Thus, the manager with direct intervention
does not need to know the utility functions of the users, $(U_1,\ldots,U_N)$,
in order to design an intervention mechanism that supports a target action profile.\footnote{Of course, if the target action profile
depends on $(U_1,\ldots,U_N)$, the manager needs to know $(U_1,\ldots,U_N)$ to determine it.}
This property can be considered as the robustness of direct intervention
with respect to the utilities of the users.

To draw a contrast, consider an alternative intervention scenario where
the manager intervenes through pricing.
In such a scenario, the action profile of the regular users determines their average data rates, i.e.,
$r_i(a) = \gamma_i a_i \prod_{j \in \mathcal{N} \setminus \{i\}} (1-a_j)$ for all
$i \in \mathcal{N}$, while a pricing mechanism
specifies the payments that the regular users make depending on their
action profile, i.e., $f(a) = (f_1(a),\ldots,f_N(a))$, where $f_i(a)$ is
the payment of user $i$. Thus, the utility function of user $i$
is given by
\begin{align*}
u_i(f(a),a) = U_i(r_i(a)) - f_i(a).
\end{align*}
As can be seen from the above expression, pricing affects utility by taking away
utility units from the users, and thus the shapes and scales of the utility functions
matter to the manager when designing optimal pricing mechanisms. For example,
consider a pricing mechanism
that charges each user an amount proportional to its average data rate,
i.e., $f_i(a) = p_i r_i(a)$ for all $i \in \mathcal{N}$, where
$p_i$ is the price of unit data rate for user $i$. Then the pricing
mechanism supports an action profile $a^*$ when the manager sets
$p_i = U'_i(r_i(a^*))$ for all $i \in \mathcal{N}$,
assuming that $U_i$ is differentiable and concave for all $i \in \mathcal{N}$.
This example illustrates that, in contrast to direct intervention, the manager needs to know the utility functions of the users, $(U_1,\ldots,U_N)$, in order to
design a pricing mechanism that supports a target action profile.

\section{Conclusion}

This paper presents the intervention framework that is aimed to optimize
the objective of the manager, or the network performance, taking the incentives of selfish users
into account. We have highlighted the generality and advantages of
our framework. In particular, we have pointed out the advantages of
intervention over pricing in terms of implementation
and informational requirement.
To facilitate analysis in the intervention framework, we have developed a new class of games
called intervention games and its solution concept called intervention equilibrium.
Our analytic results in this paper are
limited to the special case of a benevolent manager with perfect monitoring.
It is our plan for future research to investigate intervention mechanisms
in the case of imperfect monitoring, analyzing how errors in signals affect
optimal intervention mechanisms and the network performance.

\appendix[Proof of Proposition~\ref{prop:affine}]

\begin{IEEEproof}
To prove that $f_{a^*,c^*}$ supports $a^*$ is equivalent to show
\begin{align} \label{eq:maxaffine}
a_i^* \in \arg \max_{a_i \in A_i} u_i(f_{a^*,c^*}(a_i, a_{-i}^*), a_i, a_{-i}^*)
\end{align}
for all $i \in \mathcal{N}$. Note that $f_{a^*,c^*}(a_i, a_{-i}^*) = \left[ c_i^*
(a_i - a_i^*) + \underline{a}_0 \right]_{\underline{a}_0}^{\overline{a}_0}$.
We consider three cases depending on the sign of $c_i^*$.

\emph{Case 1:} $c_i^* = 0$.

By \eqref{eq:ci}, we have $\partial u_i(\underline{a}_0, a^*)/\partial a_i = 0$.
Also, we have $f_{a^*,c^*}(a_i, a_{-i}^*) = \underline{a}_0$ for all
$a_i \in A_i$. Thus, the objective function in \eqref{eq:maxaffine}
reduces to $u_i(\underline{a}_0, a_i, a_{-i}^*)$. The condition
\eqref{eq:soc1} implies that $u_i(\underline{a}_0, a_i, a_{-i}^*)$
is a concave function with respect to $a_i$ on $A_i$. Also, the first-order
optimality condition is satisfied at $a_i = a_i^*$ since
$\partial u_i(\underline{a}_0, a^*)/\partial a_i = 0$.
Therefore, $a_i^*$ maximizes $u_i(\underline{a}_0, a_i, a_{-i}^*)$
on $A_i$.

\emph{Case 2:} $c_i^* > 0$.

Since $\partial u_i(\underline{a}_0, a^*)/\partial a_0 < 0$, we have
$\partial u_i(\underline{a}_0, a^*)/\partial a_i > 0$ by \eqref{eq:ci}.
First, consider $a_i \in [\underline{a}_i, a_i^*]$. In this region,
$f_{a^*,c^*}(a_i, a_{-i}^*) = \underline{a}_0$, and thus the objective
function can be written as $u_i(\underline{a}_0, a_i, a_{-i}^*)$.
Since the condition \eqref{eq:soc21} implies that $u_i(\underline{a}_0, a_i, a_{-i}^*)$
is concave with respect to $a_i$ on $[\underline{a}_i, a_i^*]$,
$u_i(\underline{a}_0, a_i, a_{-i}^*)$ is strictly increasing in $a_i$ on
$[\underline{a}_i, a_i^*]$.

Second, consider $a_i \in [a_i^*, \min \{\overline{a}_i, a_i^* + (\overline{a}_0 - \underline{a}_0)/c_i^* \}]$.
In this region, $f_{a^*,c^*}(a_i, a_{-i}^*) = c_i^* (a_i - a_i^*) + \underline{a}_0$, and thus the objective
function can be written as $u_i(c_i^* (a_i - a_i^*) + \underline{a}_0, a_i, a_{-i}^*)$.
The first derivative of $u_i(c_i^* (a_i - a_i^*) + \underline{a}_0, a_i, a_{-i}^*)$
with respect to $a_i$ is given by
\begin{align*}
\left( c_i^* \frac{\partial u_i}{\partial a_0} +
\frac{\partial u_i}{\partial a_i} \right) \bigg|_{(a_0, a_i, a_{-i}) = (c_i^* (a_i - a_i^*) + \underline{a}_0, a_i, a_{-i}^*)},
\end{align*}
while the second derivative is given by the left-hand side of \eqref{eq:soc22}.
The first derivative is zero at $a_i = a_i^*$ by \eqref{eq:ci}, while
the second derivative is non-positive by \eqref{eq:soc22}. Hence, the first
derivative is non-positive on $(a_i^*, \min \{\overline{a}_i, a_i^* + (\overline{a}_0 - \underline{a}_0)/c_i^* \})$,
and thus $u_i(c_i^* (a_i - a_i^*) + \underline{a}_0, a_i, a_{-i}^*)$
is non-increasing in $a_i$ on $[a_i^*, \min \{\overline{a}_i, a_i^* + (\overline{a}_0 - \underline{a}_0)/c_i^* \}]$.

Lastly, consider $a_i \in [a_i^* + (\overline{a}_0 - \underline{a}_0)/c_i^*, \overline{a}_i]$.
In this region, $f_{a^*,c^*}(a_i, a_{-i}^*) = \overline{a}_0$, and thus the objective
function can be written as $u_i(\overline{a}_0, a_i, a_{-i}^*)$.
Since the first derivative of $u_i(\overline{a}_0, a_i, a_{-i}^*)$
with respect to $a_i$ is non-positive on $(a_i^* + (\overline{a}_0 - \underline{a}_0)/c_i^*, \overline{a}_i)$
by \eqref{eq:soc23}, $u_i(\overline{a}_0, a_i, a_{-i}^*)$
is non-increasing in $a_i$ on $[a_i^* + (\overline{a}_0 - \underline{a}_0)/c_i^*, \overline{a}_i]$.

\emph{Case 3:} $c_i^* < 0$.

In this case, $\partial u_i(\underline{a}_0, a^*)/\partial a_i < 0$ and
\begin{align*}
f_{a^*,c^*}(a_i, a_{-i}^*) = \left\{
\begin{array}{ll}
\overline{a}_0 \quad &\textrm{for $a_i \in [\underline{a}_i, a_i^* + (\overline{a}_0 - \underline{a}_0)/c_i^*]$,}\\
c_i^* (a_i - a_i^*) + \underline{a}_0 \quad &\textrm{for $a_i \in [\max \{\overline{a}_i,
a_i^* + (\overline{a}_0 - \underline{a}_0)/c_i^* \}, a_i^*]$,}\\
\underline{a}_0 \quad &\textrm{for $a_i \in [a_i^*, \overline{a}_i]$}.
\end{array} \right.
\end{align*}
Following an analogous argument as in Case 2, we can show that
the objective function is non-decreasing in $a_i$ on $[\underline{a}_i, a_i^*]$
and strictly decreasing on $[a_i^*, \overline{a}_i]$, implying
that $a_i = a_i^*$ maximizes the objective function on $A_i$.

Note that if the inequalities in \eqref{eq:soc1}, \eqref{eq:soc22}, and \eqref{eq:soc32}
are strict, we have $a_i = a_i^*$ as a unique maximizer for all $i \in \mathcal{N}$.
\end{IEEEproof}


\begin{thebibliography}{99}

\bibitem{garg} R. Garg, A. Kamra, and V. Khurana, ``A game-theoretic approach towards
congestion control in communication networks,'' \emph{Comput. Commun. Review},
vol. 32, no. 3, pp. 47--61, Jul. 2002.

\bibitem{jpark} J. Park and M. van der Schaar, ``Stackelberg contention games
in multiuser networks,'' {\it EURASIP J. Advances Signal Process.}, vol. 2009,
Article ID 305978, 15 pages, 2009.

\bibitem{korilis} Y. A. Korilis, A. A. Lazar, and A. Orda, ``Achieving
network optima using Stackelberg routing strategies,''
\emph{IEEE/ACM Trans. Netw.}, vol. 5, no. 1, pp. 161--173, Feb. 1997.

\bibitem{ma} R. T. Ma, V. Misra, and D. Rubenstein,
``An analysis of generalized slotted-Aloha protocols,''
\emph{IEEE/ACM Trans. Netw.}, vol. 17, no. 3, pp. 936--949, Jun. 2009.

\bibitem{yisu2} Y. Su and M. van der Schaar, ``A new perspective on multi-user
power control games in interference channels,'' \emph{IEEE
Trans. Wireless Commun.}, vol. 8, no. 6, pp. 2910--2919, Jun. 2009.

\bibitem{varian} J. K. MacKie-Mason and H. R. Varian, ``Pricing congestible
network resources,'' \emph{IEEE J. Sel. Areas Commun.}, vol. 13, no. 7, pp.
1141--1149, Sep. 1995.

\bibitem{basar} T. Ba\c sar and R. Srikant, ``A Stackelberg network game with a
large number of followers,'' \emph{J. Optimization Theory Applicat.}, vol. 115,
no. 3, pp. 479--490, Dec. 2002.

\bibitem{bloem} M. Bloem, T. Alpcan, and T. Ba\c sar, ``A Stackelberg game for
power control and channel allocation in cognitive radio networks,'' in \emph{Proc. ValueTools}, 2007.

\bibitem{low} S. H. Low and D. E. Lapsley, ``Optimization flow control---I:
Basic algorithm and convergence,''
\emph{IEEE/ACM Trans. Netw.}, vol. 7, no. 6, pp. 861--874, Dec. 1999.

\bibitem{mailath} G. Mailath and L. Samuelson, \emph{Repeated Games
and Reputations: Long-run Relationships}. Oxford, U.K.: Oxford Univ.
Press, 2006.

\bibitem{mas} A. Mas-Colell, M. D. Whinston, and J. R. Green, \emph{Microeconomic
Theory}. Oxford, U.K.: Oxford Univ. Press, 1995.

\bibitem{nisan} N. Nisan, T. Roughgarden, \'E. Tardos, and V. V. Vazirani,
Eds., \emph{Algorithmic Game Theory}. Cambridge, U.K.: Cambridge Univ. Press, 2007.

\bibitem{myerson} R. Myerson, \emph{Game Theory: Analysis of Conflict}.
Cambridge, MA: Harvard Univ. Press, 1991.

\bibitem{hamed} A. H. Mohsenian-Rad, J. Huang, M. Chiang, and V. W. S. Wong,
``Utility-optimal random access without message passing,'' \emph{IEEE
Trans. Wireless Commun.}, vol. 8, no. 3, pp. 1073--1079, Mar. 2009.

\end{thebibliography}
\end{document}